\documentclass[preprint,review,3p,12pt]{elsarticle}
\usepackage{amsmath}
\usepackage{graphics}
\usepackage{color}
\usepackage{array}
\usepackage{booktabs}
\usepackage{rotating}

\newcolumntype{P}[2]{%
  >{\begin{turn}{#1}\begin{minipage}{#2}\small\raggedright\hspace{0pt}}l%
  <{\end{minipage}\end{turn}}%
}

\journal{Physica A.}

\begin{document}

\begin{frontmatter}

\title{Solutions for a $q$--generalized Schr\"odinger equation of entangled interacting  particles}
\author[1,4]{L. G. A. Alves}
\ead{lgaalves@dfi.uem.br}
\author[1,2,4]{H. V. Ribeiro}
\author[1,4]{M. A. F. Santos} 
\author[1,4]{R. S. Mendes} 
\author[1,3,4]{E. K. Lenzi} 

\address[1]{Departamento de F\'isica, Universidade Estadual de Maring\'a - Maring\'a, PR 87020-900, Brazil}
\address[2]{Departamento de F\'isica, Universidade Tecnol\'ogica Federal do Paran\'a - Apucarana, PR 86812-460, Brazil}
\address[3]{Departamento de F\'isica, Universidade Estadual de Ponta Grossa - Ponta Grossa, PR 87030-900, Brazil}
\address[4]{National Institute of Science and Technology for Complex Systems, CNPq - Rio de Janeiro, RJ 22290-180, Brazil}

\begin{abstract}
We report on the time dependent solutions of the $q-$generalized Schr\"odinger equation proposed by Nobre \textit{et al}. [Phys. Rev. Lett. 106, 140601 (2011)]. Here we investigate the case of two free particles and also the case where two particles were subjected to a Moshinsky-like potential with time dependent coefficients. We work out analytical and numerical solutions for different values of the parameter $q$ and also show that the usual Schr\"odinger equation is recovered in the limit of $q\rightarrow 1$. An intriguing behavior was observed for $q=2$, where the wave function displays a ring-like shape, indicating a bind behavior of the particles. Differently from the results previously reported for the case of one particle, frozen states appear only for special combinations of the wave function parameters in case of $q=3$.
\end{abstract}

\end{frontmatter}
\section*{Highlights}
Solutions of a nonlinear Schr\"odinger equation;

Time dependent wave functions;

Free particles and Moshinsky-like potential.

\section{Introduction}
Tsallis statistics~\cite{Tsallis1,Curado}, based on the entropy
\begin{eqnarray}
S_{q}=k\frac{1-Tr\rho^{q}}{q-1}\,,
\end{eqnarray}
where $k$ is a constant and $q$ is a real parameter, has shown to be useful in the description of several systems where the additivity fails such as systems characterized by non--Markovian processes~\cite{non-Markovian1,non-Markovian2,grigolini,haroldo}, nonergodic dynamics and long--range many--body interactions~\cite{Long-Range Interaction}. The nonadditive characteristic of $S_{q}$~\cite{Mendes,Tsallis2,Tsallis3} has also brought new insights to several works in the area of complex systems~\cite{Tsallis2,Tsallis3} and contributed to diverse problems of quantum mechanics~\cite{Tirnakli,plastino2,Majtey,Zander,Silva,Vignat,Malacarne,Zander2,Costa,Santos,Tirnakli2}. Recently, Nobre, Rego-Monteiro and Tsallis have proposed a nonlinear Schr\"odinger equation~\cite{Nobre1} (NRT equation) that admits soliton--like solutions ($q$--plane waves) with possible applications in several areas of physics, including nonlinear optics, superconductivity, plasma physics, and darkmatter. Since the applicability of linear equations in physics is usually restricted to idealized systems~\cite{Nobre1}, the advance in the understanding of real systems makes necessary the use of nonlinear equations several times~\cite{Scott}.
For a system with particle of mass $m$, the NRT equation can be written as
\begin{equation}
i\hbar\frac{\partial }{\partial t} \left( \frac{\Phi (\vec{x},t)}{\Phi_{0}}\right)=-\frac{1}{2-q}\frac{\hbar^2}{2 m} \nabla^2 \left(\frac{\Phi (\vec{x},t)}{\Phi_{0}}\right)^{2-q}+V(\vec{x},t)\,\left(\frac{\Phi (\vec{x},t)}{\Phi_{0}}\right)^{q},
\label{edp}
\end{equation}
where the scaling constant $\Phi_{0}$ guarantees the appropriate units for the different physical terms appearing in the equation, $i$ is the imaginary unity and $\hbar$ is the Planck's constant. An interesting aspect related to this equation is the non--Markovian effect emerging from the presence of the nonlinearity, i.e., $q\neq 1$. A similar situation is found in the porous media equation \cite{PorousMedia1}, whose solutions may exhibit a short or a long tailed behavior and are connected to anomalous diffusion \cite{PorousMedia2}. This intriguing equation has been the focus of intensive research in recent
years~\cite{plastino4,plastino5,plastino3,Monteiro,Nobre2}. For instance, a family of time dependent wave packet solutions have been also investigated in Ref.~\cite{plastino2},  a quasi-stationary solution for the Moshinsky model has been proposed in Ref.~\cite{plastino3}, and a classical field-theoretic approach has been considered in Refs.~\cite{Monteiro,Nobre2}.

Here we study the problem of two particles in the context of the NRT equation focusing on time dependent solutions. We worked on the case for free particles (Section 2) and particles subjected to a time dependent Moshinsky-like potential (Section 3). For these situations, we observe that the nonlinearity creates an entanglement between the particles, which is not present in the usual scenario and it is essential for describing physical reality~\cite{plastino1} inherent to quantum mechanics~\cite{Amico,Bell}. We summarize our results and conclusions in Section 4.

\section{Free particles}

Let us start our discussion by considering the $q$--generalized Schr\"odinger equation for a system of two particles in the absence of potential by assuming $m_{1}=m_{2}=m$ and defining $\psi=\Phi/\Phi_0$ without loss of generality. The time dependent Moshinsky-like potential is discussed in the next section. For this case the NRT equation takes the following form
\begin{eqnarray}
i\hbar\frac{\partial }{\partial t} \psi (x_{1},x_{2},t)=-\frac{1}{2-q}\frac{\hbar^2}{2 m}\left(\frac{\partial^{2}}{\partial x_{1}^2 }+\frac{\partial^{2}}{\partial x_{2}^2 }\right)\psi (x_{1},x_{2},t)^{2-q}\;,
\label{eq:nrt}
\end{eqnarray}
where $x_{1}$ and $x_{2}$ represent the particles coordinates.

\begin{figure}[!h]
\centering
\includegraphics[scale=0.43]{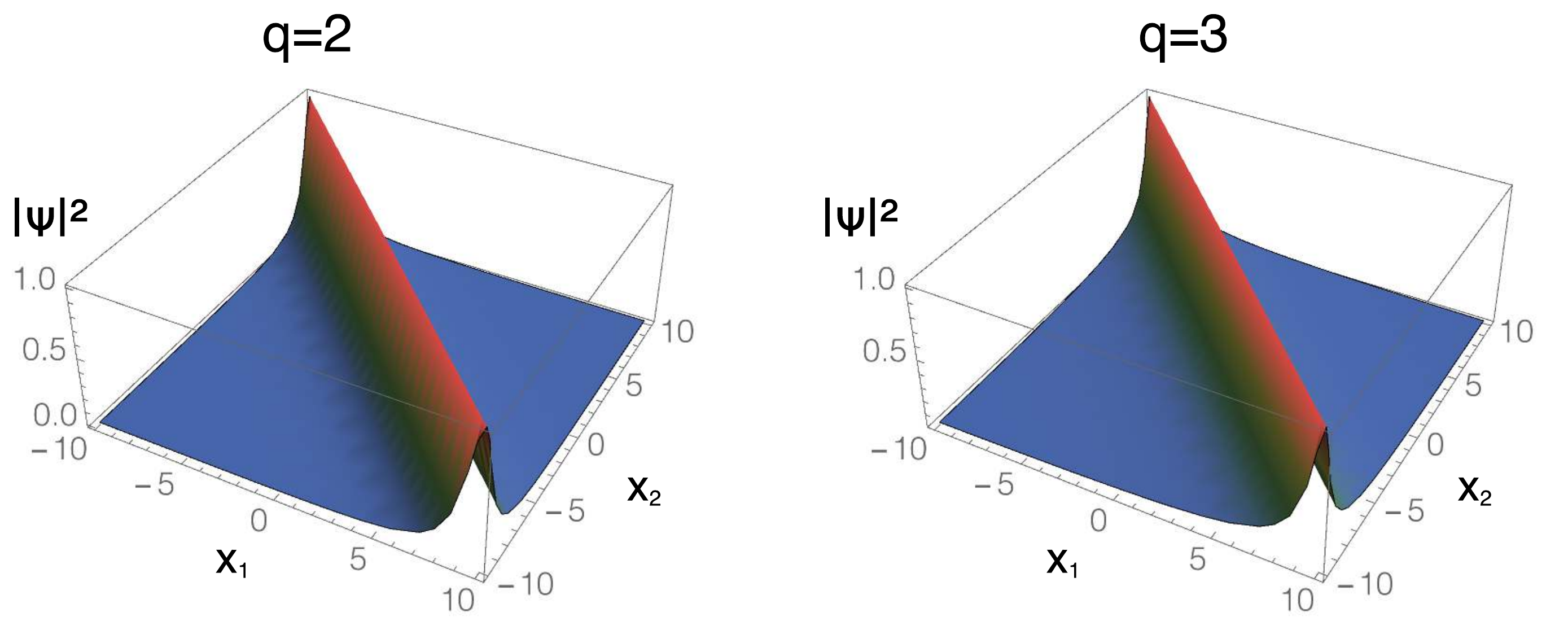}
\caption{Behavior of Eq.(\ref{qproduct}) for two different values of $q$ in order to illustrate the entanglement of the particles due to the nonlinearity of the NRT equation (Eq.(\ref{eq:nrt})).}
\label{qwavefunction}
\end{figure}

Similarly to the situation worked out in Ref.~\cite{Nobre1}, Eq.(\ref{eq:nrt}) has the $q$--plane wave as solution
\begin{eqnarray}
\label{qproduct}
\psi (x_{1},x_{2},t)=\exp_{q}\left[i(k_{1}x_{1}-\omega_{1}t)\right]\otimes_{q}\exp_{q}\left[i(k_{2}x_{2}-\omega_{2}t)\right],
\end{eqnarray}
where $\otimes_{q}$ is the $q$--product~\cite{Produto}. This result shows that the nonlinearity of the Eq.(\ref{eq:nrt}) entangled the particles, leading us to different behaviors
dependent on $q$ (see Fig.(\ref{qwavefunction})). 
{ It is worth noting that $|\psi|^2$ obtained from the NRT equation does not has a probabilistic interpretation for $q \neq 1$. In order to recover that probabilistic interpretation, we could follow the procedures proposed by Rego-Monteiro and Nobre~\cite{Monteiro}, and introduce a second field defined by means of an additional non-linear equation coupled with the NRT equation for $q \neq 1$. Furthermore, recovering the probabilistic interpretation of the NRT equation under the action of an spatial and time dependent potential (such as the Moshinsky-like potential worked out in the next section), in contrast with the free particle case, still is an open problem. Even in the free particle case, to obtain general time-dependent solutions for the NRT equation represents a cumbersome task due to the intrinsic complexity of $\psi$ that still need to be incorporated in this second equation for obtaining the second field. For this reason, we have focused our study only on the solutions for the NRT equation, that is, on $\psi$. Although this simplified approach lacks an exact  probabilistic interpretation, for $q$ not so far from one we could expected that $|\psi|^2$ will keep some information on the probability of find particles in a given position and time; also, this analysis can provides new insights and a better understating of the NRT equation.}

In order to explicit solve Eq.(\ref{eq:nrt}), we use the $q$--Gaussian package as an ansatz, 
\begin{eqnarray}
\label{eq:ansatz}
\psi(x_{1},x_{2},t)=\left[1-(1-q)\left(a(t)\,x_{1}^2 + b(t)\,x_{2}^2+ c(t)\,x_{1}\,x_{2}+d(t)\right)\right]^{\frac{1}{1-q}}
\end{eqnarray}
where $a$, $b$, $c$, and $d$ are appropriate time dependent coefficients to be found. The presence of a crossing term in Eq.(\ref{eq:ansatz}) covers a more general scenario characterized by an initial situation of mixing between the variables $x_{1}$ and $x_{2}$. 

By substituting (\ref{eq:ansatz}) in (\ref{eq:nrt}), the left and right sides of the NRT equation becomes,
\begin{eqnarray}
i\hbar\frac{\partial }{\partial t} \psi (x_{1},x_{2},t)=-i \hbar  \left(x_{1}^2 a'(t)+x_{2}^2 b'(t)+x_{1}\, x_{2}\, c'(t)+d'(t)\right) \psi(x_{1},x_{2},t)^{q},
\end{eqnarray}
and
\begin{eqnarray}
\lefteqn{-\frac{1}{2-q}\frac{\hbar^2}{2 m}\left(\frac{\partial^{2}}{\partial x_{1}^2 }+\frac{\partial^{2}}{\partial x_{2}^2 }\right)\psi (x_{1},x_{2},t)^{2-q}/{\left(\frac{\hbar^2}{2 m} \psi(x_{1},x_{2},t)^q\right)}= } \nonumber \\
 &&{2 a(t) \left((q-1) b(t) \left(x_{1}^2+x_{2}^2\right)+(q-3) x_{1} x_{2} c(t)+(q-1) d(t)+1\right)+2 (q-3) x_{1}^2 a(t)^2+} \nonumber \\
 &&{2 b(t) ((q-3) x_{1} x_{2} c(t)+(q-1) d(t)+1)+2 (q-3) x_{2}^2 b(t)^2-c(t)^2 \left(x_{1}^2+x_{2}^2\right)}.
\end{eqnarray}
Thus, Eq.(\ref{eq:ansatz}) satisfies the NRT equation with $a$, $b$, $c$ and $d$ being solution of the following set of coupled ordinary differential equations:
\begin{eqnarray}
a'(t)&=&\frac{i \hbar }{2\, m} \left(2\, a(t) ((q-3)\, a(t)+(q-1)\, b(t))-c(t)^2\right),\\
b'(t)&=&\frac{i \hbar }{2\, m}  \left(2\, b(t) ((q-1)\, a(t)+(q-3)\, b(t))-c(t)^2\right),\\
c'(t)&=&\frac{i \hbar }{m} (q-3)\,   c(t)\, (a(t)+b(t)),\\
d'(t)&=&\frac{i \hbar}{m}  (a(t)+b(t))\, ((q-1)\, d(t)+1).
\label{edos}
\end{eqnarray}

Note that there is a relationship between these parameters. By considering the difference $a'(t)-b'(t)$, we obtain
\begin{eqnarray}
a'(t)-b'(t)&=&\frac{c'(t)}{c(t)}\,(a(t)-b(t)).
\end{eqnarray}
Thus, we can write the following relationship for $a$, $b$ and $c$:
\begin{eqnarray}
a(t)=b(t)+\kappa \,c(t),
\label{eq:parametersrelation}
\end{eqnarray}
where $\kappa=(a(0)-b(0))/c(0)$.

{ A general analytical solution for this set of equations with arbitrary initial conditions is a hard task; however, note that any initial state with $c(0)=0$ lead us to $c(t)=$constant for all later times. Likewise, any initial state with $a(0)=b(0)$ will preserve this symmetry for all later times (see Eq.(\ref{eq:parametersrelation})). Also, if $c(0)=0$ and $a(0)=b(0)$, we have the following solutions:
\begin{eqnarray}
a(t)&=& -\frac{c_0}{2 \sqrt{q-2}} \tanh \left(\frac{i\, \hbar\,  c_0\,  t\,}{m}\sqrt{q-2}\, +2\, c_0\, c_1 \sqrt{q-2}\right),\\
b(t)&=&a(t),\\
c(t)&=&c_0,\\
d(t)&=&\frac{1}{1-q}+c_2\, e^{\frac{2\, i\, \hbar\,(q-1)\, t  }{m}} \left[ \cos \left(\frac{ \hbar\,  c_0\,  t\,}{m}\sqrt{q-2}\, +2\,i\, c_0\, c_1 \sqrt{q-2}\right) \right]^{-\frac{q-1}{2 (q-2)}},
\end{eqnarray}
where, $c_0$, $c_1$ and $c_2$ are constant that depend on the initial conditions. Naturally, it is possible to obtain other numerical and analytical solutions for $a$, $b$, $c$ and $d$ to investigate the behavior of $|\psi(x_1,x_2,t)|^2$ for the cases that we shall present.}

\subsection{Limit case $q\rightarrow 1$}

\begin{figure}[!h]
\centering
\includegraphics[scale=0.34]{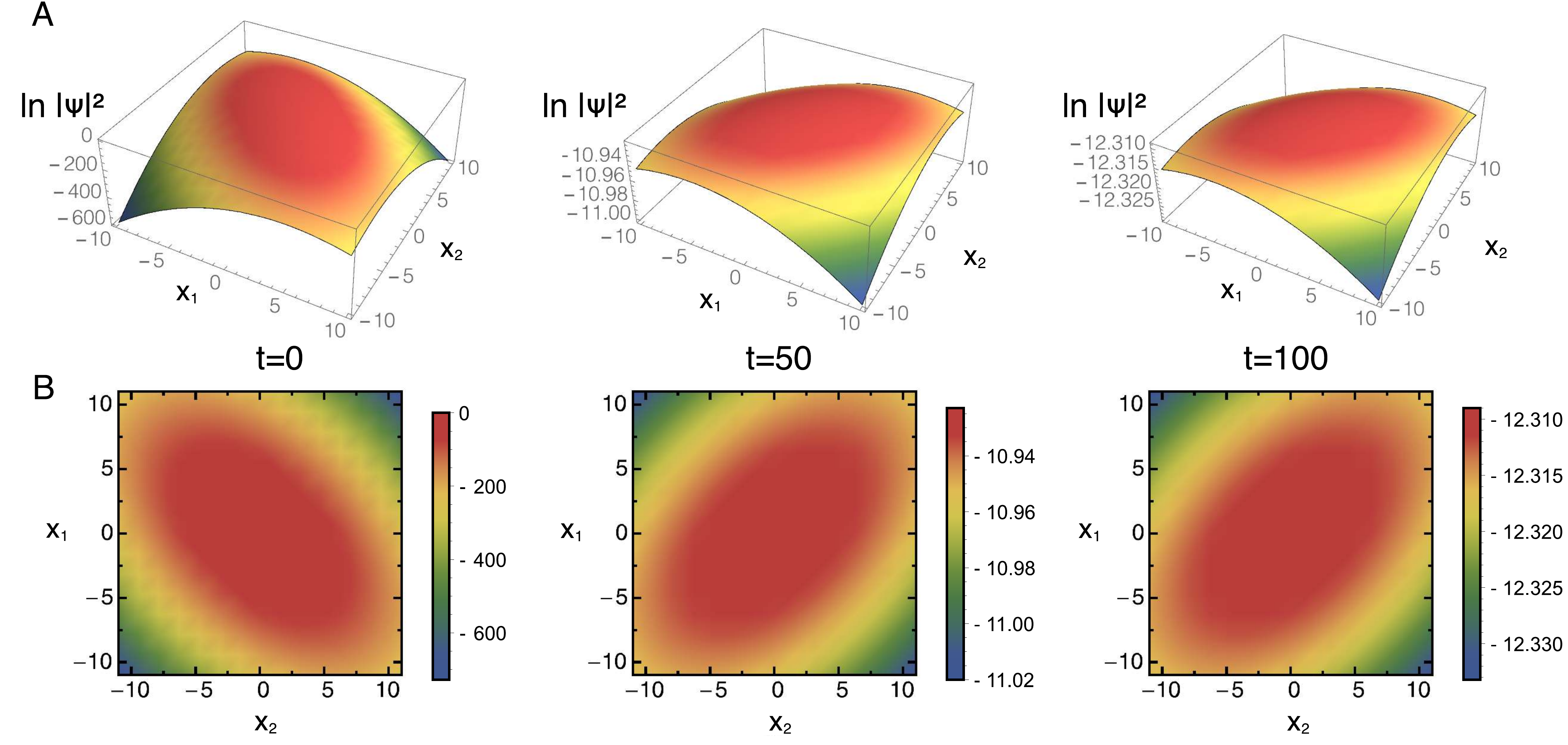}
\caption{Evolution of the $\mid \psi(x_{1},x_{2},t) \mid^2$ for the entangled system of free particles in the linear limit case, $q=1$. Here, we have used $m=1$, $\hbar=1$, and the following initial conditions: $a(0)=1$, $b(0)=1$, $c(0)=1$ and $d(0)=1$. We have employed the $\ln{\mid \psi(x_{1},x_{2},t) \mid^2}$ in order to smooth the surfaces. Figure (A) shows a 3D-plot and figure (B) shows a density-plot of the $\ln{\mid \psi(x_{1},x_{2},t) \mid^2}$ in function of the positions $x_1$ and $x_2$ of the two particles.}
\label{fig1}
\end{figure}

Let us discuss the case that recovers the linear Schr\"odinger equation for a systems of two entangled free particles. In this case, the set of differential equations for $a$, $b$, $c$ and $d$ is
\begin{eqnarray}
a'(t)&=&-\frac{i \hbar }{m} \left(2\, a(t)^2+\frac{c(t)^2}{2}\right),\\
b'(t)&=&-\frac{i \hbar }{m} \left(2\, b(t)^2+\frac{c(t)^2}{2}\right),\\
c'(t)&=&-\frac{2\,i \hbar }{m}\, c(t)\, \left(a(t)\, + b(t)\, \right),\\
d'(t)&=&\frac{i \hbar }{m}\,\left(a(t)+ b(t)\right).
\end{eqnarray}
Considering the difference $a'(t)-b'(t)$, Eq.(~\ref{eq:parametersrelation}), and using this result in the above set of equations, we can rewrite $c'(t)$ as
\begin{eqnarray}
c'(t)=-\frac{\,i \hbar }{m}\,\left(4\, b(t)\,c(t)+ 2\,\kappa\,c(t)^2\right).
\label{eq:c1}
\end{eqnarray}
If we sum $b'(t)$ and $c'(t)$ we have that,
\begin{eqnarray}
b'(t)+c'(t)&=&-\frac{\,i \hbar }{m}\left(b(t)^2+2\,b(t)\,c(t)+\left(\kappa+\frac{1}{4}\right)\,c(t)^2\right).
\label{eq:sum1}
\end{eqnarray}
Notice that $\kappa$ is a free parameter that depends only on the initial conditions of $a$, $b$ and $c$. For simplicity, we choose $\kappa=3/4$, yielding
\begin{eqnarray}
b'(t)+c'(t)&=&-\frac{2\,i \hbar }{m}\left(b(t)+c(t)\right)^2
\end{eqnarray}
and, therefore
\begin{eqnarray}
b(t)=\frac{m}{2\,i\,\hbar\, t}+c(t).
\end{eqnarray}
We solve the set of equations for the parameters $a$, $b$, $c$ and $d$, leading to:
\begin{eqnarray}
a(t)&=&\frac{2\, m\, \hbar +i\, \lambda\, m^2\, t}{-2\, \lambda\, m\, t^2\, \hbar +11\, i\, \hbar ^2\,t},\\
b(t)&=&\frac{m}{2\,i\,\hbar\,t}+\frac{2 m}{t \left(2\, \lambda\, m\, t-11\, i\, \hbar \right)},\\
c(t)&=&\frac{2\, m}{t \left(2\, \lambda\, m\, t-11\, i\, \hbar \right)},\\
d(t)&=&\frac{11\, \hbar +4\, i\, \lambda\, m\, t}{22\, t\, \hbar +4\, i\, \lambda\, m\, t^2},
\end{eqnarray}
where $\lambda$ depends on the initial conditions. It is worth noting that when $t\rightarrow \infty$, we have $a(t)\sim b(t) \sim d(t) \sim 1/t$ and $c(t) \sim 1/t^{2}$. We have also evaluated {\text{a}} numerical solution by considering the initial conditions $a(0)=1$, $b(0)=1$, $c(0)=1$, $d(0)=1$ and plotted the evolution of $\mid \psi(x_{1},x_{2},t) \mid^2$ for different times in Fig.(\ref{fig1}). In this case, the contribution of the coupling term, $c(t)$, vanishes faster than the others for long times, making the shape of the wave function similar to the case of one free particle~\cite{plastino2}. 

\subsection{Evolution of the wave function for q=2}

\begin{figure}[!h]
\centering
\includegraphics[scale=0.34]{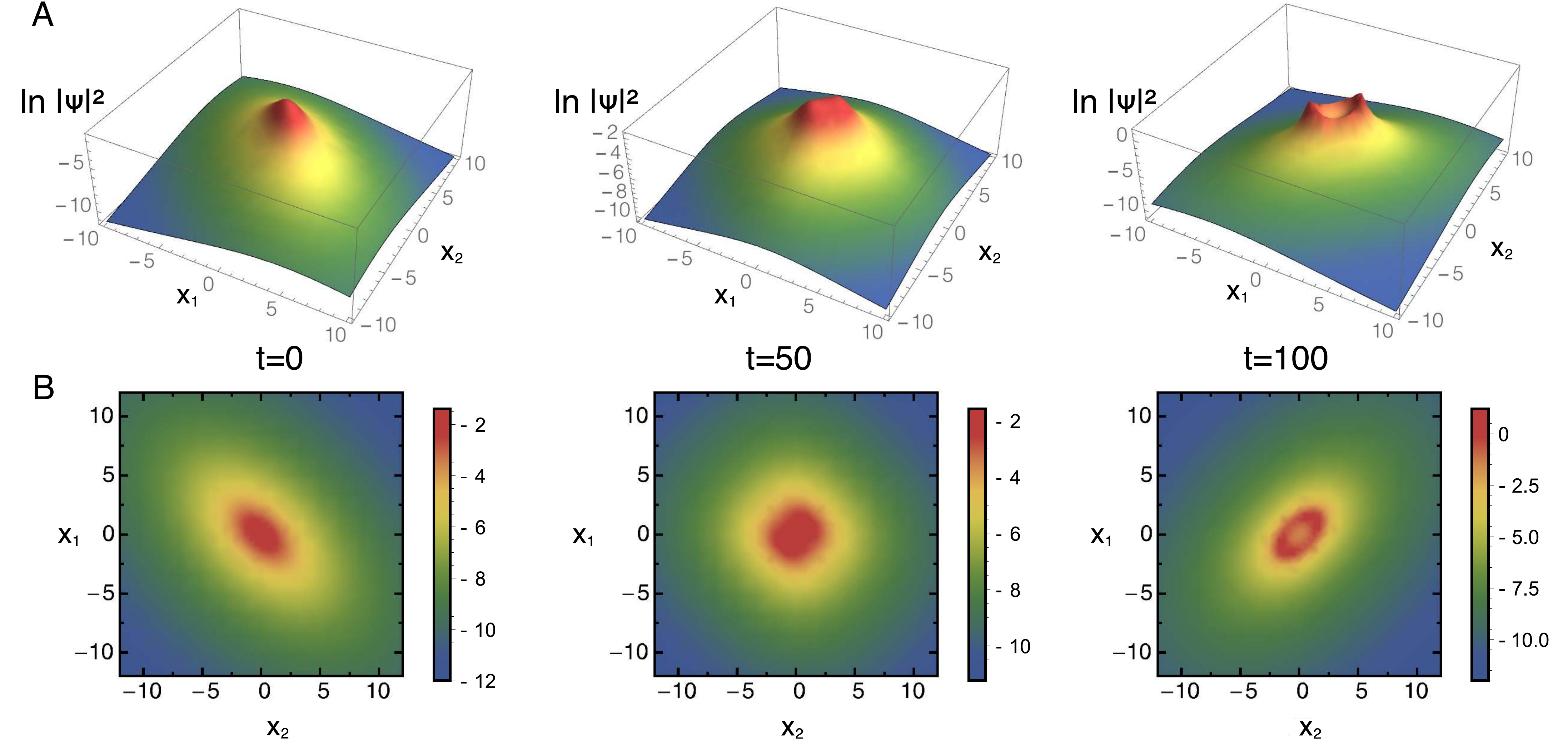}
\caption{Evolution of the $\mid \psi(x_{1},x_{2},t) \mid^2$ for the entangled system of free particles in the case $q=2$. Here, we have used $m=1$, $\hbar=1$, and the following initial conditions: $a(0)=1$, $b(0)=1$, $c(0)=1$ and $d(0)=1$. We have employed the $\ln{\mid \psi(x_{1},x_{2},t) \mid^2}$ in order to smooth the surfaces. Figure (A) shows a 3D-plot and figure (B) shows a density-plot of the $\ln{\mid \psi(x_{1},x_{2},t) \mid^2}$ in function of the positions $x_1$ and $x_2$ of the two particles.}
\label{fig2}
\end{figure}

For the nonlinear case $q=2$, the set of differential equations takes the form:
\begin{eqnarray}
a'(t)&=& \frac{i \hbar }{2\,m} \left(2\, a(t) (b(t)-a(t))-c(t)^2\right),\\
b'(t)&=& \frac{i \hbar }{2\,m}\left(2\, b(t)\, (a(t)-b(t))-  c(t)^2\right),\\
c'(t)&=& -\frac{i \hbar }{m}\, c(t)\, (a(t)+b(t)),\\
d'(t)&=& \frac{i \hbar }{m}\, (d(t)+1)\, (a(t)+b(t)).
\end{eqnarray}
Because the inherent complexity of this set of equations, we assume that $a(t)=b(t)$ in order to simplify the problem. Under this assumption, we have the following solutions:
\begin{eqnarray}
a(t)&=&b(t)= \pm \sqrt{\frac{c_1}{2}} \tan \left(\sqrt{2\,c_1}\, c_2\,-\frac{i \sqrt{2\,c_1}\, t\, \hbar }{m}\right),\\
c(t)&=& \mp\sqrt{2c_1}\sec\left(\sqrt{2\,c_1}\, c_2-\frac{i \sqrt{2\,c_1}\, t\, \hbar }{m}\right),\\
d(t)&=& -1+c_3 \cosh \left(\frac{\sqrt{2\,c_1} }{m}\left(t\, \hbar +i\, c_2\, m\right)\right),
\end{eqnarray}
where $c_1$, $c_2$ and $c_3$ are constants that depend on the initial conditions. In particular, the constant $c_{1}$ can be determined via $c_{1}=2a^{2}(0)-c^{2}(0)/2$ and the others by using the previous results. In Fig.(\ref{fig2}), we show a numerical solution for $\mid \psi(x,y,t) \mid^2$ where we observe the emergence of two peaks around $t=50$, which evolve to a ring-like shape in $t=100$ (see Fig.(\ref{fig2}B)) and bind the $\psi$ function of the two particles. This behavior was also found in the case of one particle~\cite{plastino2}.

\subsection{Analytical solution for q=3 and the absence of ``frozen" solutions}

\begin{figure}[!h]
\centering
\includegraphics[scale=0.34]{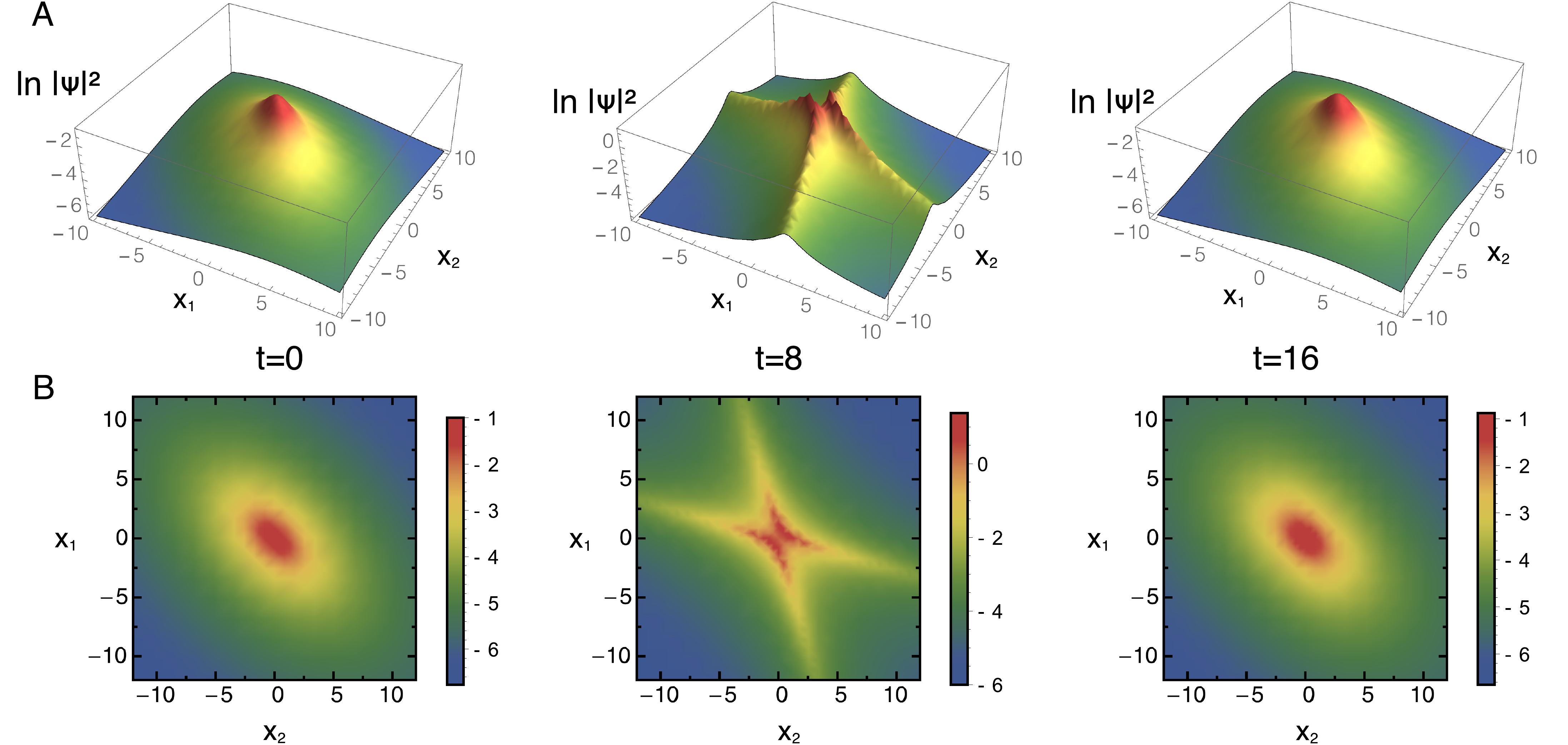}
\caption{Evolution of the $\mid \psi(x_{1},x_{2},t) \mid^2$ for the entangled system of free particles in the case $q=3$. Here, we have used $m=1$, $\hbar=1$, $a_0=1$, $c_0=1$ and $d_0=1$. We have employed the $\ln{\mid \psi(x_{1},x_{2},t) \mid^2}$ in order to smooth the surfaces. Figure (A) shows a 3D-plot and figure (B) shows a density-plot of the $\ln{\mid \psi(x_{1},x_{2},t) \mid^2}$ in function of the positions $x_{1}$ and $x_{2}$ of the two particles.}
\label{fig3}
\end{figure}

\begin{figure}[!h]
\centering
\includegraphics[scale=0.43]{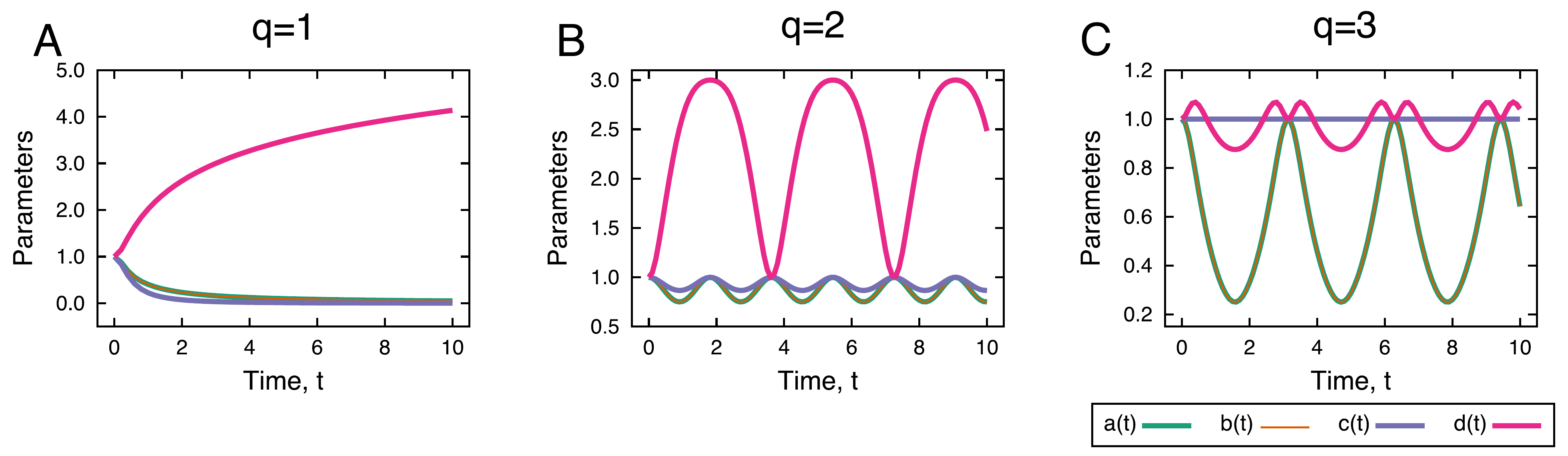}
\caption{Evolution of the time dependent parameters. The plots (A) $q=1$ and (B) $q=2$ are numerical solutions, while (C) is the analytical solutions found for the case $q=3$, Eqs. (37), (38) and (39). Note that $a(t)=b(t)$ due to the choice of the initial conditions.}
\label{parameters1}
\end{figure}

For the case of $q=3$, the set of differential equations for the parameters $a$, $b$, $c$ and $d$ can be reduced to
\begin{eqnarray}
a'(t)&=&\frac{i\,\hbar }{2\, m }  \left(4\, a(t)\, b(t)-c(t)^2\right),\\
b'(t)&=&\frac{i\,\hbar }{2\, m  } \left(4\, a(t)\, b(t)-c(t)^2\right),\\
c'(t)&=&0,\\
d'(t)&=&\frac{i\, \hbar }{m } (2\, d(t)+1)\, (a(t)+b(t)).
\end{eqnarray}
Notice that $a'(t)=b'(t)$ and the solution for $c'(t)$ is straightforward. Also, we have that $a(t)=b(t)$ by imposing the same initial conditions for Eqs.(33) and (34). Under this assumption, the time dependent parameters are given by
\begin{eqnarray}
a(t)&=&b(t)=-\frac{c_{0}}{2}\left(\frac{1-{{\cal{A}}}e^{-i\,\alpha\,c_0\, t}}{1+{{\cal{A}}}e^{-i\,\alpha\,c_0\, t}}\right),\\
c(t)&=&c_0,\\
d(t)&=&\frac{1}{2}\left(1+2d_0\right)e^{2i\alpha\int_{0}^{t}a(t')dt'}-\frac{1}{2},
\end{eqnarray}
where $a_0$, $c_0$ and $d_0$ depend on the initial conditions, ${\cal{A}}=(c_0/2+a_0)/(c_0/2-a_0)$ and $\alpha=2\hbar/m$ (see Fig.~\ref{fig3}).
Curilef \textit{et al.}~\citep{plastino2} have showed the existence of frozen solutions for $q=3$ in the case of one particle. In our case, there are no frozen states for $\psi(x_{1},x_{2},t)$ due to the interactions between the particles caused by the nonlinearity of the Eq.(\ref{eq:nrt}). We illustrate this aspect by plotting the evolution of the parameters  $a$, $b$, $c$ and $d$ (in absolute values) for $q=1$, $q=2$ and $q=3$, as shown in Fig.(\ref{parameters1}). The frozen states appears only in special cases where $4\, a(t)\, b(t)=c(t)^2$ and $d(t)=0$. We further observe that if $a_0=-c_0/2$, then $a(t)=b(t)=-c_0/2$ and $d(t)\sim e^{-i \alpha c_0 t}$, which implies in a quasi-stationary behavior { for  $\psi(x_{1},x_{2},t)$ }. In this scenario, the solution ``pulsates'' with a period $\frac{2 \pi}{\alpha c_0}$ and for $d_0=0$ we obtain another frozen state, similarly to the case of one free particle~\cite{plastino2}.  

\section{Moshinsky-like potential}

Entanglement features of atomic systems are often studied by considering the Moshinsky model~\cite{Moshinsky,Moshinsky2}. This model has been intensively investigated in the recent years as a test for validating several approximations in atomic and molecular physics. In this context, we investigate the solutions of the NRT equation for a system of two particles subjected to a Moshinsky-like potential, i.e.,
\begin{equation}
V(x_{1},x_{2},t)=\alpha(t)\,x_{1}^2+\beta(t)\,x_{2}^2+\gamma(t)\,x_{1}\,x_{2}\,+\eta(t)
\label{potential}
\end{equation}
where $\alpha$,  $\beta$, $\gamma$ e $\eta$ are time dependent parameters.

Considering the potential (\ref{potential}) and by applying the same procedures of Section 2, the ansatz (\ref{eq:ansatz}) is solution of the NRT equation if the time dependent parameters satisfy the following set of equations:
\begin{eqnarray}
a'(t)&=&\frac{i}{2\, m\, \hbar} \left(\hbar ^2\, \left(2\, a(t) ((q-3)\, a(t)+(q-1)\, b(t))-c(t)^2\right)+2\, m\, \alpha (t)\right),\\
b'(t)&=&\frac{i}{2\, m\, \hbar} \left(\hbar ^2\, \left(2\, b(t) ((q-1)\, a(t)+(q-3)\, b(t))-c(t)^2\right)+2\, m\, \beta (t)\right),\\
c'(t)&=&\frac{i}{m\, \hbar}\left((q-3)\, \hbar ^2\, c(t)\, (a(t)+b(t))+m\, \gamma (t)\right),\\
d'(t)&=&\frac{i}{m\,\hbar} \left(\hbar ^2\, (a(t)+b(t))\, ((q-1)\, d(t)+1)+m\, \eta (t)\right).
\label{edosp}
\end{eqnarray}

Naturally, we recover the free particle case if $\alpha=\beta=\gamma=\eta=0$. We study numerical solutions for this set of equations since analytical approaches represent a cumbersome problem. We have focused on numerical solutions for $\psi (x_{1},x_{2},t)$ considering particular values of $q$, $\alpha(t)=\beta(t)=\eta(t)=1$ and an exponential decay for the coupling factor, this is,  $\gamma(t)=e^{-t}$. This latter hypothesis makes the coupling between the particles strong for small times and weak for long times. 

\subsection{Limit case $q\rightarrow1$}
\begin{figure}[!h]
\centering
\includegraphics[scale=0.34]{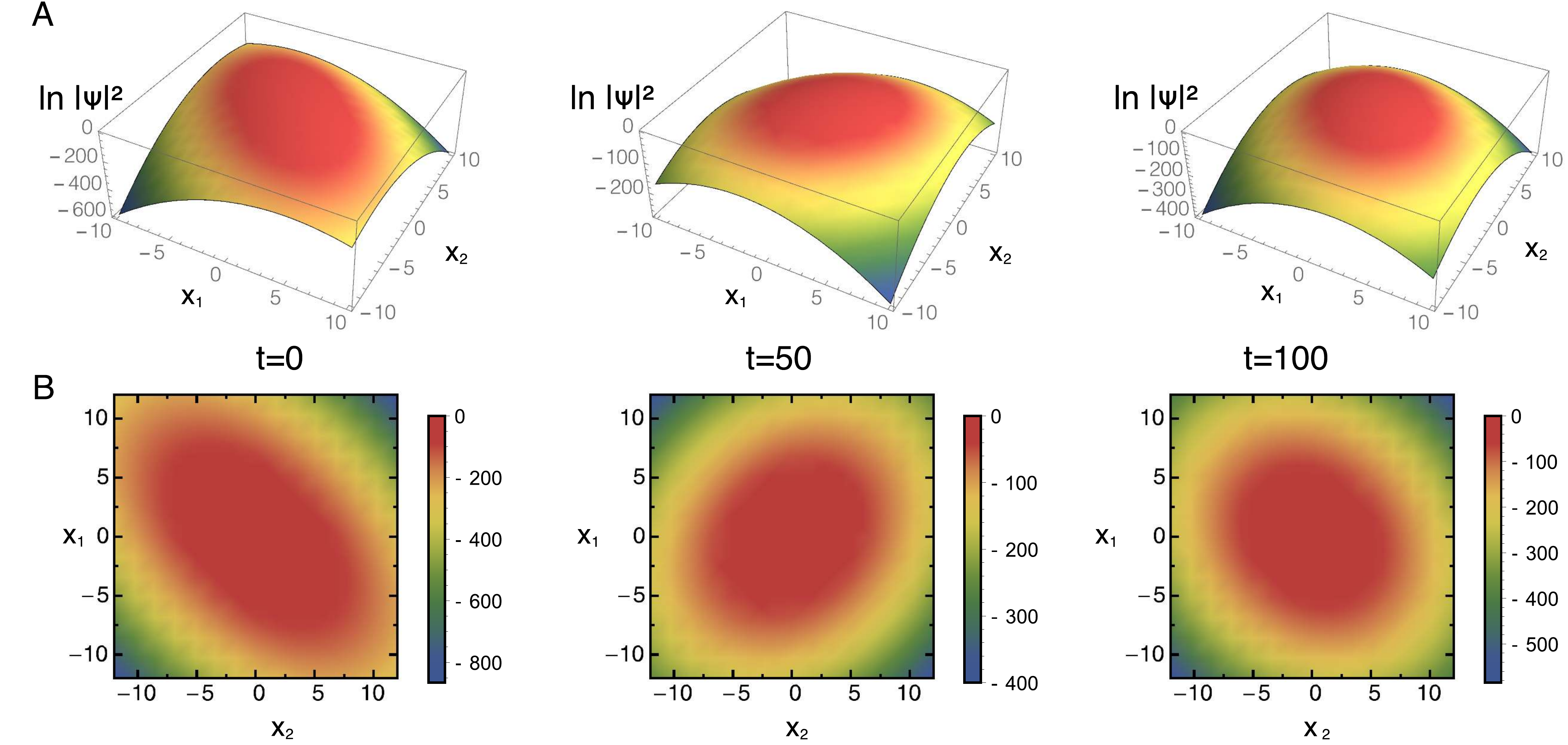}
\caption{Evolution of the $\mid \psi(x_{1},x_{2},t) \mid^2$ for the entangled system subject to a Monshinsky-like potential in the limit case of $q=1$. Here, we have used $m=1$, $\hbar=1$, $a(0)=1$, $b(0)=1$, $c(0)=1$, $d(0)=1$, and the following functions for the Moshinsky-like potential: $\alpha(t)=1$, $\beta(t)=1$, $\gamma(t)=e^{-t}$, $\eta(t)=1$. We have employed the $\ln{\mid \psi(x_{1},x_{2},t) \mid^2}$ in order to smooth the surfaces. Figure (A) shows a 3D-plot and figure (B) shows a density-plot of the $\ln{\mid \psi(x_{1},x_{2},t) \mid^2}$ in function of the positions $x_1$ and $x_2$ of the two particles.}
\label{fig4}
\end{figure}

For $q=1$ we recover the linear Schr\"odinger equation and the set of differential equations  becomes:
\begin{eqnarray}
a'(t)&=&\frac{i}{2\, m\, \hbar }\left(\hbar ^2 \left(-4 a(t)^2-c(t)^2\right)+2\, m\, \alpha (t)\right),\\
b'(t)&=&\frac{i}{2\, m\, \hbar }\left(\hbar ^2 \left(-4 b(t)^2-c(t)^2\right)+2\, m\, \beta (t)\right),\\
c'(t)&=&\frac{i}{m\, \hbar } \left(-2 \hbar ^2 c(t)\, (a(t)+b(t))+m\, \gamma (t)\right),\\
d'(t)&=&\frac{i}{m\, \hbar } \left(\hbar ^2 (a(t)+b(t))+m\, \eta (t)\right).
\end{eqnarray}
The solution for this case is shown in Fig.~\ref{fig4} where we note that the shape of the wave
functions are visually similar to the case of two free particles (see Fig.(\ref{fig2})). 

\subsection{Evolution of the wave function for q=2}
\begin{figure}[!h]
\centering
\includegraphics[scale=0.34]{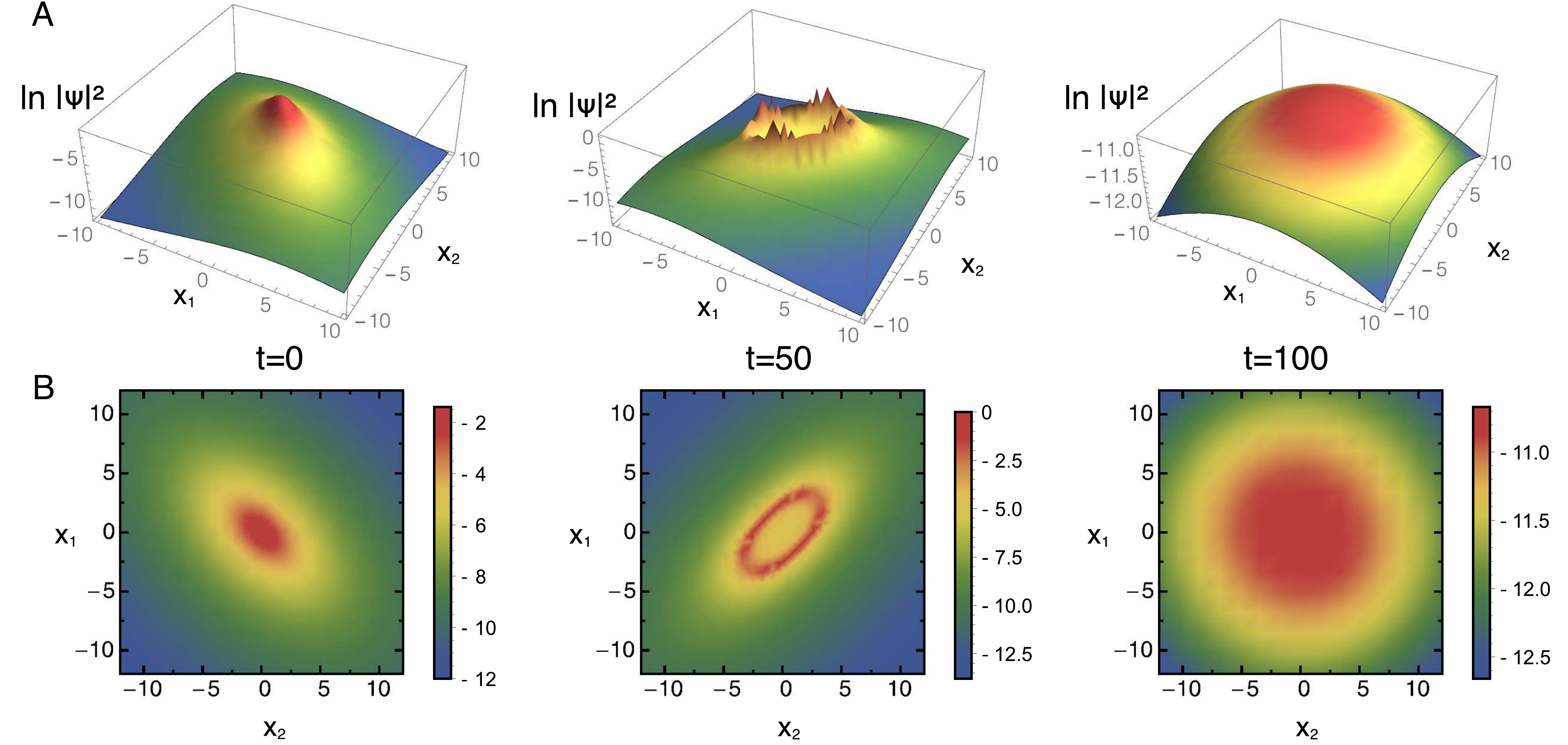}
\caption{Evolution of the $\mid \psi(x_{1},x_{2},t) \mid^2$ for the entangled systems subject to a Monshinsky-like potential for $q=2$. Here, we have used $m=1$, $\hbar=1$, $a(0)=1$, $b(0)=1$, $c(0)=1$, $d(0)=1$, and the following functions for the Moshinsky-like potential: $\alpha(t)=1$, $\beta(t)=1$, $\gamma(t)=e^{-t}$, $\eta(t)=1$. We have employed the $\ln{\mid \psi(x_{1},x_{2},t) \mid^2}$ in order to smooth the surfaces. Figure (A) shows a 3D-plot and figure (B) shows a density-plot of the $\ln{\mid \psi(x_{1},x_{2},t) \mid^2}$ in function of the positions $x_{1}$ and $x_{2}$ of the two particles.}
\label{fig5}
\end{figure}

In this section, we show a numerical solution for the nonlinear Schr\"odinger equation with $q=2$. In this case, the set of differential equations are:
\begin{eqnarray}
a'(t)&=&\frac{i}{2 m \hbar} \left(\hbar^2 \left(2 a(t) (b(t)-a(t))-c(t)^2\right)+2 m \alpha (t)\right),\\
b'(t)&=&\frac{i}{2 m \hbar} \left(\hbar^2 \left(2 b(t) (a(t)-b(t))-c(t)^2\right)+2 m \beta (t)\right)\\
c'(t)&=&\frac{i}{m \hbar } \left(-\hbar^2 c(t) (a(t)+b(t)+m \gamma (t))\right),\\
d'(t)&=&\frac{i}{m \hbar } \left(\hbar^2 (d(t)+1) (a(t)+b(t))+m \eta (t)\right).
\end{eqnarray}
and the numerical solutions for $\psi(x_{1},x_{2},t)$ are shown in Fig.(\ref{fig5}). due to the interaction with the potential, the binding behavior appears in different times, see for example $t=50$. We further note that the ring-like form is deformed and that $\mid \psi(x_{1},x_{2},t) \mid^2$  exhibit some peaks.

\subsection{Evolution of the wave function for q=3 under the Moshinsky-like potential and the absence of ``frozen" solutions}
\begin{figure}[!h]
\centering
\includegraphics[scale=0.34]{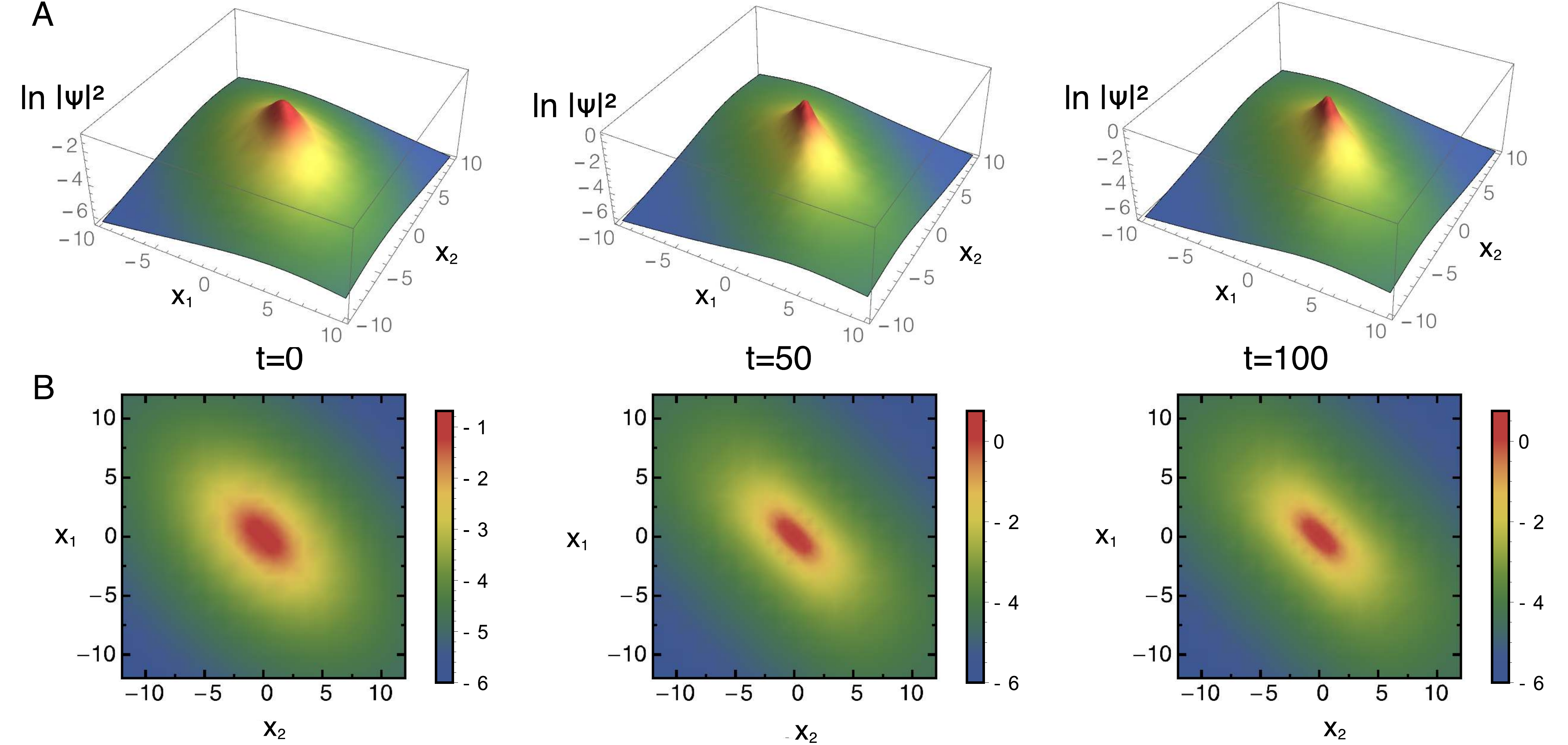}
\caption{Evolution of the $\mid \psi(x_{1},x_{2},t) \mid^2$ for the entangled systems subject to a Monshinsky-like potential for $q=3$. Here, we have used $m=1$, $\hbar=1$, $a(0)=1$, $b(0)=1$, $c(0)=1$, $d(0)=1$, and the following functions for the Moshinsky-like potential: $\alpha(t)=1$, $\beta(t)=1$, $\gamma(t)=e^{-t}$, $\eta(t)=1$. We have employed the $\ln{\mid \psi(x,y,t) \mid^2}$ in order to smooth the surfaces. Figure (A) shows a 3D-plot and figure (B) shows a density-plot of the $\ln{\mid \psi(x_{1},x_{2},t) \mid^2}$ in function of the positions $x_1$ and $x_2$ of the two particles.}
\label{fig6}
\end{figure}

\begin{figure}[!h]
\centering
\includegraphics[scale=0.43]{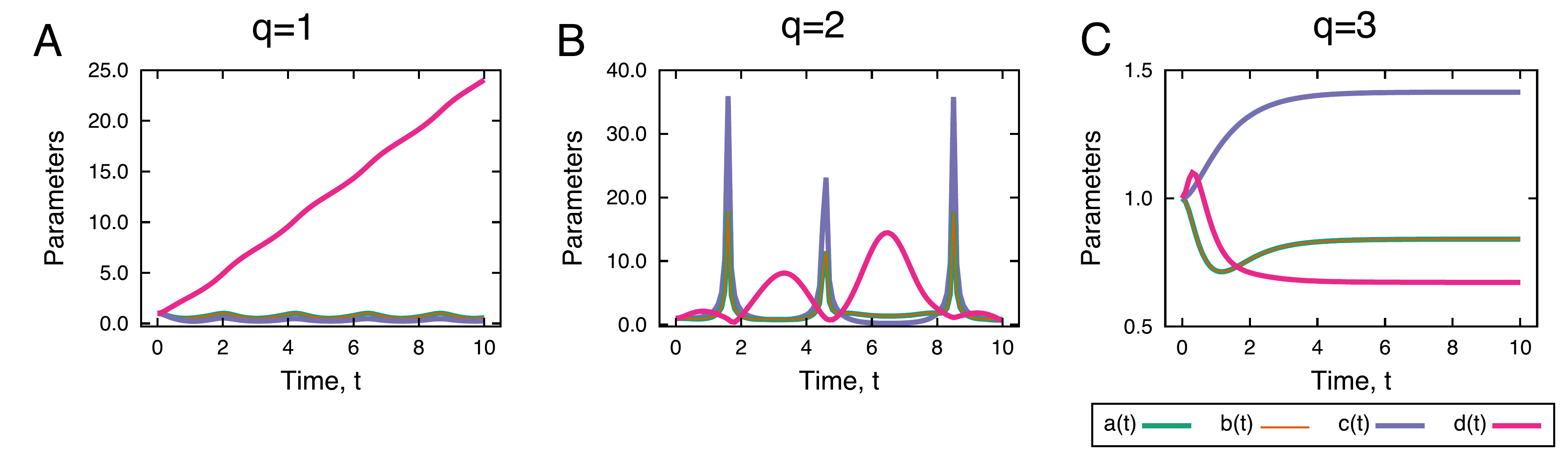}
\caption{Evolution of the time dependent parameters. The plots (A) $q=1$ and (B) $q=2$ (C) $q=3$ are the numerical solutions. Note that $a(t)=b(t)$ due to the choice of the initial conditions.}
\label{parameters2}
\end{figure}

We also investigate the existence of frozen solutions considering $q=3$ under Moshinsky-like potential (\ref{potential}). For this case, the set of differential equations are:
\begin{eqnarray}
a'(t)&=&\frac{i}{2\, m\, \hbar } \left(\hbar ^2 \left(4\, a(t)\, b(t)-c(t)^2\right)+2\, m\, \alpha (t)\right),\\
b'(t)&=&\frac{i}{2\, m\, \hbar } \left(\hbar ^2 \left(4\, a(t)\, b(t)-c(t)^2\right)+2\, m\, \beta (t)\right),\\
c'(t)&=&\frac{i\, \gamma (t)}{\hbar },\\
d'(t)&=&\frac{i}{m\, \hbar } \left(\hbar ^2\, (2\, d(t)+1)\, (a(t)+b(t))+m\, \eta (t)\right).
\end{eqnarray}
Note that all the  differential equations are time dependent. Even  for the particular case where $4\, a(t)\, b(t)=c(t)^2$ and $d(t)=0$ the parameters are time dependent, indicating that the Moshinsky-like potential prohibits the existence of frozen solutions. Figure~\ref{fig6} illustrates  $|\psi|^2$ for this case and Fig.~\ref{parameters2} displays the evolution of the parameters $a$, $b$, $c$ and $d$ for $q=1$, $q=2$ and $q=3$. Notice that for $q=3$, the parameters approach constant values over time, achieving a stationary behavior.

\section{Summary and conclusions}
We found time dependent solutions of the nonlinear Schr\"odinger equation proposed by Nobre \textit{et al}. in [Phys. Rev. Lett. 106, 140601 (2011)]. These solutions are obtained in terms of $q$--exponentials. We have investigated the case of free particles and also the case where the particles were subjected to a Moshinsky-like potential with time dependent coefficients. We showed that the solution of the usual Schr\"odinger equation is recovered for $q\rightarrow 1$. We also investigated solutions for $q\neq1$ where we observe a bind behavior of $ \psi$ for the case $q=2$ and the absence of frozen states for $q=3$, differently from the results previously reported for the case of one particle.

\section{Acknowledgements}
This work has been supported by the CNPq, CAPES and Funda\c{c}\~ao Arauc\'aria (Brazilian agencies).

\end{document}